\documentclass{elsart4-1}

 \usepackage{graphicx}

\usepackage{amssymb}

\usepackage[english]{babel}

\newtheorem{e-proposition}[theorem]{Proposition}

\newtheorem{e-definition}[theorem]{Definition\rm}

\def\og{\leavevmode\raise.3ex\hbox{$\scriptscriptstyle\langle\!\langle$~}}
\def\fg{\leavevmode\raise.3ex\hbox{~$\!\scriptscriptstyle\,\rangle\!\rangle$}}

\begin{document}

\centerline{Physics }
\begin{frontmatter}


\selectlanguage{english}
\title{Spin-polarized proximity effect in superconducting junctions}


\selectlanguage{english}
\author T. Yokoyama and 
\ead{h042224m@mbox.nagoya-u.ac.jp}
\author Y. Tanaka
\ead{ytanaka@nuap.nagoya-u.ac.jp}
\address{Department of Applied Physics, Nagoya University, Nagoya, 464-8603, Japan
\\
and CREST, Japan Science and Technology Corporation (JST) Nagoya, 464-8603,
Japan}


\medskip

\begin{abstract}
We study spin dependent phonomena in superconducting junctions in both 
ballistic and diffusive regimes. 
For ballistic junctions we study both ferromagnet / $s$- and 
$d$-wave superconductor junctions and  two dimensional electron gas / $s$-wave 
superconductor junctions with  Rashba spin-orbit coupling. It is shown that the exchange field alway suppresses the conductance while the Rashba spin-orbit coupling can enhance it. 
In the latter part of the article we study  the diffusive
ferromagnet / insulator / $s$- and $d$-wave superconductor (DF/I/S) junctions, where the proximity effect can be enhanced by the exchange field in contrast to common belief. 
This resonant proximity effect in these 
junctions is studied for various situations:  Conductance of the
junction and density of states of the DF are calculated by changing the
heights of the insulating barriers at the interfaces, the magnitudes of the
resistance in DF,  the exchange field in DF, the
transparencies of the insulating barriers and the angle between the normal
to the interface and the crystal axis of $d$-wave superconductors $\alpha$.
It is shown that the resonant proximity effect originating from the exchange
field in DF strongly influences the tunneling conductance and density of
states. We clarify the followings: for $s$-wave junctions,  a sharp zero bias conductance peak (ZBCP) appears due to the
resonant proximity effect. The magnitude of this ZBCP can exceed its value in 
normal states in contrast to the one observed in diffusive normal metal / 
superconductor junctions. We find  similar structures  to the conductance in the density of states. For $d$-wave junctions at $\alpha=0$, we also find a result  similar to that in $s$-wave junctions. 
The magnitude of the resonant ZBCP at $\alpha=0$ can exceed the one at 
$\alpha/\pi=0.25$ due to the formation of the mid gap Andreev resonant states. 
\vskip 0.5\baselineskip



}
\end{abstract}
\end{frontmatter}


\selectlanguage{english}

\section{Introduction}
In normal metal / supercunductor (N/S) junctions, a unique scattering process occurs in low energy transport: Andreev reflection (AR)\cite{Andreev}. The AR is a process that  an electron injected from N with energy below the energy gap $\Delta$ is  converted into a reflected hole. Taking the AR into account,  Blonder, Tinkham and Klapwijk (BTK) proposed the formula for the  calculation of the tunneling conductance\cite{BTK}. It revealed the gap like structure or the doubling of tunnelilng conductance due to the AR. This method was extended to   normal metal / unconventional superconductor (N/US) junctions\cite{TK95}. It is shown 
that a zero bias conductance peak  (ZBCP) appears when the mid gap Andreev resonant state (MARS) is formed due to the anisotropy of  US. 

The BTK theory was also extended to ferromagnet / superconductor (F/S) or ferromagnet / unconventional superconductor (F/US) junctions\cite{FS} and used to estimate the spin polarization of the F layer experimentally
\cite{Tedrow,Upadhyay,Soulen}. In F/S junctions, AR is suppressed because the retro-reflectivity is broken by  the spin-polarization in the F 
layer\cite{de Jong}. To clarify spin dependent transport phenomena 
is important to fabricate a new device manipulating electron's spin. 
Nowadays, there are many  works about charge transport of electrons relevant to electron's spin. 

Among recent works, many efforts have been devoted to study the effect of spin-orbit coupling on transport properties of two dimensional electron gas (2DEG)\cite{Hirsch,Streda,Schliemann,Sinova}. 
The pioneering work by Datta and Das suggested the way to control the precession of the spins of electrons by the Rashba spin-orbit coupling (RSOC)\cite{Rashba} in F/2DEG/F junctions\cite{Datta}. This spin-orbit coupling depends on the 
applied field and can be tuned by a gate voltage. It also  gives the 
off-diagonal elements of  the velocity operator\cite{Molenkamp}.  
There are several works about  spin dependent transport in the presence of 
RSOC  \cite{Edelstein,Inoue}. 

\begin{figure}[htb]
\begin{center}
\scalebox{0.4}{
\includegraphics[width=28.0cm,clip]{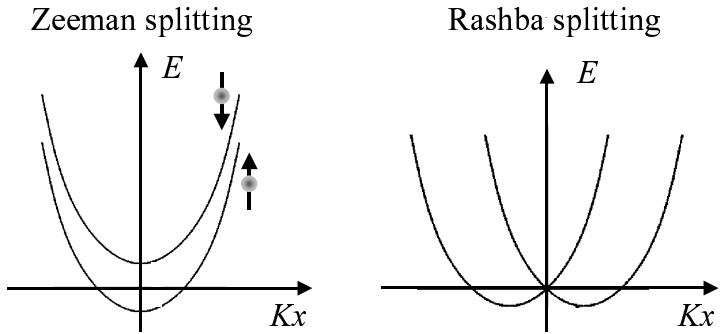}}
\end{center}
\par
\caption{ Zeeman vs. Rashba splitting.}
\label{f0}
\end{figure}

 The RSOC induces an energy splitting, but the  energy splitting doesn't break 
the time reversal symmetry unlike an exchange splitting in ferromagnet.  Therefore transport properties in 2DEG/S junctions may be qualitatively different from  those in F/S junctions. As far as we know, in 2DEG/S junctions  the effect of RSOC on transport phenomena is not studied well. Recent experimental and theoretical advances in spintronics stimulate us to challange this problem. We illustrate the two kind of splittigs in Fig. \ref{f0}.

The first purpose of this article is to calculate the tunneling conductance in F/S and 2DEG/S junctions and clarify how the exchange field and the RSOC affect 
it.  We think the obtained results are useful  for a better understanding of related experiments in mesoscopic F/S and 2DEG/S junctions.

 On the other hand, in diffusive junctions the physics is clearly different from that in ballistic junctions.  In diffusive normal
metal / superconductor (DN/S) junctions, proximity effect  plays an
important role in the low energy transport. The phase coherence between
incoming electrons and Andreev reflected holes persists in DN at a
mesoscopic length scale and results in strong interference effects on the
probability of AR \cite{Hekking}. 
One of the striking experimental manifestations is the zero bias conductance
peak (ZBCP) \cite%
{Giazotto,Klapwijk,Kastalsky,Nguyen,Wees,Nitta,Bakker,Xiong,Magnee,Kutch,Poirier}%
.

A quasiclassical Green's function theory\cite{Eilenberger,Eliashberg,Larkin}
is often applied to the charge transport in DN/S junctions. Volkov, Zaitsev
and Klapwijk (VZK) solved the Usadel equations \cite{Usadel}, and showed
that this ZBCP is due to the enhancement of the pair amplitude in DN by the
proximity effect \cite{Volkov}. VZK applied the Kupriyanov and Lukichev (KL)
boundary condition for the Keldysh-Nambu Green's function \cite{KL}.
Stimulated by the VZK theory, several authors studied the charge transport
in various junctions \cite%
{Nazarov1,Yip,Stoof,Reentrance,Golubov,Takayanagi,Bezuglyi,Seviour,Belzig}.

%
Recently one of the authors \cite{TGK} developed the VZK theory for $s$-wave
superconductors using more general boundary conditions provided by the
circuit theory of Nazarov \cite{Nazarov2}. 
The boundary conditions coincide with the KL conditions when a connector is
diffusive with low transparent coefficients, while the BTK theory \cite%
{BTK} is reproduced in the ballistic regime. The extended VZK theory \cite%
{TGK} produced a crossover from a ZBCP to a zero bias conductance dip
(ZBCD). These phenomena are relevant for the actual junctions in which the
barrier transparency is not necessarily small. 

The formation of the MARS at the interface
of unconventional superconductors \cite{Buch,TK95,Kashi00,Experiments} also
generates the ZBCP as mentioned above. The generalized VZK theory was recently applied  to
unconventional superconducting junctions \cite{Nazarov3,Golubov2,p-wave}.
The formation of the MARS is naturally taken into account in this approach.
It was demonstrated that the formation of MARS in DN/$d$-wave superconductor
(DN/D) junctions strongly competes with the proximity effect.

 Above
theories treat spin independent phenomena in diffusive junctions. Calculations of tunneling
conductance in the presence of the magnetic impurities in DN /S junctions
were performed in Ref\cite{Yip2,Yoko,Volkov}. Spin dependent transport is
also realized in ferromagnet / superconductor junctions.

In diffusive ferromagnet / superconductor (DF/S) junctions Cooper pairs
penetrate into the DF layer from the S layer and have a nonzero momentum by
the exchange field\cite{Buzdin1982,Buzdin1991,Demler}. This property
produces many interesting phenomena\cite%
{Ryazanov,Kontos1,Blum,Sellier,Strunk,Radovic,Tagirov,Fominov,Rusanov,Ryazanov1,Kadigrobov,Seviour2,Leadbeater,Bergeret,Kadigrobov2}%
. One interesting consequence of the oscillations of the pair amplitude is
the spatially damped oscillating behavior of the density of states (DOS) in
a ferromagnet predicted theoretically \cite{Buzdin,Zareyan,Baladie,Bergeret2}. In the ferromagnet the exchange field usually breaks the induced Cooper
pairs. But for a weak exchange field the pair amplitude can be \textit{%
enhanced} and the energy dependent DOS can have a zero-energy peak\cite%
{Golubov3}. The DOS has been studied extensively\cite%
{Zareyan,Golubov3,Krivoruchko,Kontos} but the condition for the appearance
of the DOS peak was not studied systematically. We studied the conditions
for the appearance of such anomaly, i.e., strong enhancement of the
proximity effect and found two conditions corresponding to weak proximity
effect and strong one\cite{Yoko2}. Since DOS is a fundamental quantity, this 
\textit{resonant proximity effect} can influence various transport
phenomena.

 Another purpose of the present article  is to study the influence of the
resonant proximity effect by the exchange field on the tunneling conductance and the DOS 
in DF/ $s$- and $d$-wave superconductor junctions with Nazarov's boundary
conditions.  Weak exchange field is realized in recent experiments
with, e.g., Ni doped Pd\cite{Kontos} or magnetic semiconductor. Thus our
results may be observed in experiments.  In the latter part of this article we  calculate 
the tunneling conductance and the density of states in normal metal /
insulator / diffusive ferromagnet / insulator / $s$- and $d$-wave
superconductor (N/I/DF/I/S) junctions for various parameters such as the
heights of the insulating barriers at the interfaces, resistance $R_{d}$ in
DF, the exchange field $h$ in DF, the Thouless energy $E_{Th}$ in DF, the
transparencies of the insulating barriers and the angle between the normal
to the interface and the crystal axis of $d$-wave superconductors $\alpha$.
Throughout the paper we confine ourselves to zero temperature.

The organization of this paper is as follows. In sections II and III, we will provide
the detailed derivation of the expression for the normalized tunneling
conductance and  the results of calculations are presented for
various types of junctions, in ballistic  and diffusive junctions respectively. In section IV, the summary of the obtained
results is given.

\section{Ballistic junctions}
We consider F/S and F/$d$-wave
superconductor (F/D)  junctions. We use the same method as in Ref. \cite{FS} and the same notations.  In the following $\uparrow (\downarrow)$ denotes majority (minority) spin.
  The F/US interface 
located at $x=0$ (the $y$-axis) has an infinitely
narrow insulating barrier described by the delta function $U(x)=H\delta
(x)$. As a model of the ferromagnet we apply the Stoner model with the 
exchange potential $U$. The pair potential matrix we consider is given by
\begin{equation}
\mathord{\buildrel{\lower3pt\hbox{$\scriptscriptstyle\frown$}} 
\over \Delta } \left( \theta  \right) = \left( {\begin{array}{*{20}c}
   0 & {\Delta _{ \uparrow  \downarrow } \left( \theta  \right)}  \\
   {\Delta _{ \downarrow  \uparrow } \left( \theta  \right)} & 0  \\
\end{array}} \right)
\end{equation}
where $\theta$ denotes the direction of motions of quasiparticles measured from  the normal to the interface. 
Below we consider $s$- and $d$-wave superconductors. The pair potentials  are given by $\Delta_{\downarrow  \uparrow}=-\Delta_{\uparrow \downarrow}=\Delta$ for  $s$-wave superconductors and  $\Delta_{\downarrow  \uparrow}=-\Delta_{\uparrow \downarrow}=\Delta\cos[2(\theta - \alpha)]$ for $d$-wave superconductors where $\alpha$ denotes
the angle between the normal to the interface and the crystal axis of $d$
-wave superconductors. Here $\Delta$ denotes the energy gap. 

Applying the BTK method\cite{BTK,FS}, we obtain 
the conductance $\sigma _{S \uparrow (\downarrow) }$   for up (down) spin quasiparticle represented in the form:
\begin{eqnarray}
 \sigma _{S \uparrow }  = \sigma _{N \uparrow } \frac{{1 - \left| {\Gamma _ +  \Gamma _ -  } \right|^2 \left( {1 - \sigma _{N \downarrow } } \right) + \sigma _{N \downarrow } \left| {\Gamma _ +  } \right|^2 }}{{\left| {1 - \Gamma _ +  \Gamma _ -  \sqrt {1 - \sigma _{N \uparrow } } \sqrt {1 - \sigma _{N \downarrow } } \exp \left[ {i\left( {\varphi _ \downarrow   - \varphi _ \uparrow  } \right)} \right]} \right|^2 }}\Theta \left( {\theta _C  - \left| \theta  \right|} \right) \nonumber \\ 
  + \sigma _{N \uparrow } \frac{{1 - \left| {\Gamma _ +  \Gamma _ -  } \right|^2 }}{{\left| {1 - \Gamma _ +  \Gamma _ -  \sqrt {1 - \sigma _{N \uparrow } } \exp \left[ {i\left( {\varphi _ \downarrow   - \varphi _ \uparrow  } \right)} \right]} \right|^2 }}\Theta \left( {\left| \theta  \right| - \theta _C } \right) \\
 \sigma _{S \downarrow }  = \sigma _{N \downarrow } \frac{{1 - \left| {\Gamma _ +  \Gamma _ -  } \right|^2 \left( {1 - \sigma _{N \uparrow } } \right) + \sigma _{N \uparrow } \left| {\Gamma _ +  } \right|^2 }}{{\left| {1 - \Gamma _ +  \Gamma _ -  \sqrt {1 - \sigma _{N \uparrow } } \sqrt {1 - \sigma _{N \downarrow } } \exp \left[ {i\left( {\varphi _ \uparrow   - \varphi _ \downarrow  } \right)} \right]} \right|^2 }}\Theta \left( {\theta _C  - \left| \theta  \right|} \right) 
\end{eqnarray}
\begin{eqnarray}
 \exp \left( {i\varphi _ \downarrow  } \right) = \frac{{1 - \lambda _ -   + iZ_\theta  }}{{\sqrt {1 - \sigma _{N \downarrow } } \left( {1 + \lambda _ -   - iZ_\theta  } \right)}},\quad \exp \left( { - i\varphi _ \uparrow  } \right) = \frac{{1 - \lambda _ +   - iZ_\theta  }}{{\sqrt {1 - \sigma _{N_ \uparrow  } } \left( {1 + \lambda _ +   + iZ_\theta  } \right)}}, \\ 
 \Gamma _ \pm   = \frac{{\Delta _ \pm  \left( \theta  \right)}}{{E + \sqrt {E^2  - \left| {\Delta _ \pm  \left( \theta  \right)} \right|^2 } }}, 
\end{eqnarray}

$Z_\theta   = \frac{Z}{{\cos \theta }}$, $Z = \frac{{2mH}}{{\hbar ^2 k_F }}$ and $\theta _C  =\cos^{-1}  \sqrt {\frac{U}{{E_F }}} $ with quasiparticle energy $E$, effective mass $m$, Fermi wavenumber $k_F$ and Fermi energy $E_F$. In the above $\Theta(x)$ is the Heaviside step function and  $\sigma _{N \uparrow (\downarrow) }$ denotes the conductance for up (down) spin quasiparticle in the normal state:
\begin{equation}
\sigma _{N \uparrow }  = \frac{{4\lambda _ +  }}{{\left( {1 + \lambda _ +  } \right)^2  + Z_\theta ^2 }},\quad \sigma _{N \downarrow }  = \frac{{4\lambda _ -  }}{{\left( {1 + \lambda _ -  } \right)^2  + Z_\theta ^2 }}\Theta \left( {\theta _C  - \left| \theta  \right|} \right)
\end{equation}%

with $\lambda _ \pm   = \sqrt {1 \pm \frac{U}{{E_F \cos ^2 \theta }}} $.

Normalized conductance  is expressed as 
\begin{equation}
\sigma _T  = \frac{{\int_{ - \frac{\pi }{2}}^{\frac{\pi }{2}} {d\theta \cos \theta \left( {\sigma _{S \uparrow }  + \sigma _{S \downarrow } } \right)} }}{{\int_{ - \frac{\pi }{2}}^{\frac{\pi }{2}} {d\theta \cos \theta \left( {\sigma _{N \uparrow }  + \sigma _{N \downarrow } } \right)} }}.
\end{equation}

Next we consider a ballistic 2DEG/S junctions.
  The 2DEG/S interface 
located at $x=0$ (along the $y$-axis) has an infinitely
narrow insulating barrier described by the delta function $U(x)=H\delta
(x)$. 
The effective Hamiltonian  with RSOC is given by
\begin{equation}
H = \left( {\begin{array}{*{20}c}
   {\xi _k } & {i\lambda k_ -  \Theta \left( { - x} \right)} & 0 & {\Delta \Theta \left( x \right)}  \\
   { - i\lambda k_ +  \Theta \left( { - x} \right)} & {\xi _k } & { - \Delta \Theta \left( x \right)} & 0  \\
   0 & { - \Delta \Theta \left( x \right)} & { - \xi _k } & { - i\lambda k_ +  \Theta \left( { - x} \right)}  \\
   {\Delta \Theta \left( x \right)} & 0 & {i\lambda k_ -  \Theta \left( { - x} \right)} & { - \xi _k }  \\
\end{array}} \right)
\end{equation}
with $k_ \pm   = k_x  \pm ik_y $, $\xi _k  = \frac{{\hbar ^2 }}{{2m}}\left( {k^2  - k_F^2 } \right)$,  Fermi wave number $k_F$, Rashba coupling constant $\lambda$.

Velocity operator in the $x$-direction is given by\cite{Molenkamp}
\begin{equation}
v_x  = \frac{{\partial H}}{{\hbar \partial k_x }} = \left( {\begin{array}{*{20}c}
   {\frac{\hbar }{{mi}}\frac{\partial }{{\partial x}}} & {\frac{{i\lambda }}{\hbar }\Theta \left( { - x} \right)} & 0 & 0  \\
   { - \frac{{i\lambda }}{\hbar }\Theta \left( { - x} \right)} & {\frac{\hbar }{{mi}}\frac{\partial }{{\partial x}}} & 0 & 0  \\
   0 & 0 & { - \frac{\hbar }{{mi}}\frac{\partial }{{\partial x}}} & { - \frac{{i\lambda }}{\hbar }\Theta \left( { - x} \right)}  \\
   0 & 0 & {\frac{{i\lambda }}{\hbar }\Theta \left( { - x} \right)} & { - \frac{\hbar }{{mi}}\frac{\partial }{{\partial x}}}  \\
\end{array}} \right).
\end{equation}

Wave function $\psi(x)$  for $x \le 0$ (2DEG region) is represented using eigenfunctions of the Hamiltonian: 
\begin{equation}
\begin{array} {l}
 e^{ik_y y} \left[ {\frac{1}{{\sqrt 2 }}e^{ik_{1(2)} \cos \theta _{1(2)} x} \left( {\begin{array}{*{20}c}
   {\left(  -  \right)i\frac{{k_{1(2) - } }}{{k_{1(2)} }}}  \\
   1  \\
   0  \\
   0  \\
\end{array}} \right) + \frac{{a_{1(2)} }}{{\sqrt 2 }}e^{ik_1 \cos \theta _1 x} \left( {\begin{array}{*{20}c}
   0  \\
   0  \\
   {i\frac{{k_{1 + } }}{{k_1 }}}  \\
   1  \\
\end{array}} \right) + \frac{{b_{1(2)} }}{{\sqrt 2 }}e^{ik_2 \cos \theta _2 x} \left( {\begin{array}{*{20}c}
   0  \\
   0  \\
   { - i\frac{{k_{2 + } }}{{k_2 }}}  \\
   1  \\
\end{array}} \right)} \right. \\ 
 \left. { + \frac{{c_{1(2)} }}{{\sqrt 2 }}e^{ - ik_1 \cos \theta _1 x} \left( {\begin{array}{*{20}c}
   { - i\frac{{k_{1 + } }}{{k_1 }}}  \\
   1  \\
   0  \\
   0  \\
\end{array}} \right) + \frac{{d_{1(2)} }}{{\sqrt 2 }}e^{ - ik_2 \cos \theta _2 x} \left( {\begin{array}{*{20}c}
   {i\frac{{k_{2 + } }}{{k_2 }}}  \\
   1  \\
   0  \\
   0  \\
\end{array}} \right)} \right] 
\end{array}
\end{equation}
for an injection wave with wave number $k_{1(2)}$ where 
 $ k_1  =  - \frac{{m\lambda }}{{\hbar ^2 }} + \sqrt {\left( {\frac{{m\lambda }}{{\hbar ^2 }}} \right)^2  + k_F^2 } $,
$ k_2  =   \frac{{m\lambda }}{{\hbar ^2 }} + \sqrt {\left( {\frac{{m\lambda }}{{\hbar ^2 }}} \right)^2  + k_F^2 } $ and $k_{1(2) \pm }  = k_{1(2)} e^{ \pm i\theta _{1(2)} } $. $a_{1(2)}$ and $ b_{1(2)}$ are AR coefficients. $c_{1(2)}$ and $d_{1(2)}$ are normal reflection coefficients.
 $\theta _{1(2)}$ is an angle of the wave with wave number $k_{1(2)}$ with respect to the interface normal.

Similarly for $x \ge 0$ $\psi(x)$ (S region) is given by the linear combination of the eigenfunctions. Note that since the translational symmetry holds for the $y$-direction, the momenta parallel to the interface are conserved: $k_y=k_F \sin \theta  = k_1 \sin \theta _1  = k_2 \sin \theta _2 $.

 The wave function follows the boundary conditions\cite{Molenkamp}:
\begin{eqnarray}
 \left. {\psi \left( x \right)} \right|_{x =  + 0}  = \left. {\psi \left( x \right)} \right|_{x =  - 0}  \\ 
 \left. {v_x \psi \left( x \right)} \right|_{x =  + 0}  - \left. {v_x \psi \left( x \right)} \right|_{x =  - 0}  = \frac{\hbar }{{mi}}\frac{{2mU}}{{\hbar ^2 }}\left( {\begin{array}{*{20}c}
   1 & 0 & 0 & 0  \\
   0 & 1 & 0 & 0  \\
   0 & 0 & { - 1} & 0  \\
   0 & 0 & 0 & { - 1}  \\
\end{array}} \right)\psi \left( 0 \right) 
\end{eqnarray}

Applying BTK theory to our calculation, we obtain the dimensionless conductance  represented in the form:
\begin{equation}
\begin{array}{l}
 \sigma _s  = N_1 \int_{ - \theta _C }^{\theta _C } {\frac{1}{2}\left[ {K_{21}  + \left| {a_1 } \right|^2 K_{21}  + \left| {b_1 } \right|^2 K_{12} \lambda _{21}  - \left| {c_1 } \right|^2 K_{21}  - \left| {d_1 } \right|^2 K_{12} \lambda _{21} } \right]} \cos \theta d\theta  \\ 
  + N_2 \int_{ - \frac{\pi }{2}}^{\frac{\pi }{2}} {{\mathop{\rm Re}\nolimits} \frac{1}{2}\left[ {K_{12}  + \left| {a_2 } \right|^2 K_{21} \lambda _{12}  + \left| {b_2 } \right|^2 K_{12}  - \left| {c_2 } \right|^2 K_{21} \lambda _{12}  - \left| {d_2} \right|^2 K_{12} } \right]} \cos \theta d\theta  \\ 
 \end{array}
\end{equation}%
with 
$
K_{12}  = 1 + \frac{{k_1 }}{{k_2 }}, K_{21}  = 1 + \frac{{k_2 }}{{k_1 }}
$ and
\begin{equation}
\lambda _{12}  = \frac{{k_1 \cos \theta _1 }}{{k_2 \cos \theta _2 }} \quad \quad \lambda _{21}  = \frac{{k_2 \cos \theta _2 }}{{k_1 \cos \theta _1 }} \quad \quad
N_1  =\frac{1}{{1 + \frac{{m\lambda }}{{\hbar ^2 k_1 }}}} \quad \quad N_2  = \frac{1}{{1 - \frac{{m\lambda }}{{\hbar ^2 k_2 }}}}.
\end{equation}%
$N_1$ and $N_2$ are normalized density of states for wave number $k_1$ and $k_2$ respectively. 
The critical angle $\theta _C$ is defined as $\cos \theta _C  = \sqrt {\frac{{2m\lambda }}{{\hbar ^2 k_1 }}} $.

$\sigma _N$ is given by the conductance for normal states, i.e., $\sigma _S$ for $\Delta=0$. We define normalized conductance as $\sigma _T =\sigma _S /\sigma _N$ and a parameter $\beta$ as
$\beta  = \frac{{2m\lambda }}{{\hbar ^2 k_F }}$.

\begin{figure}[htb]
\begin{center}
\scalebox{0.4}{
\includegraphics[width=25.0cm,clip]{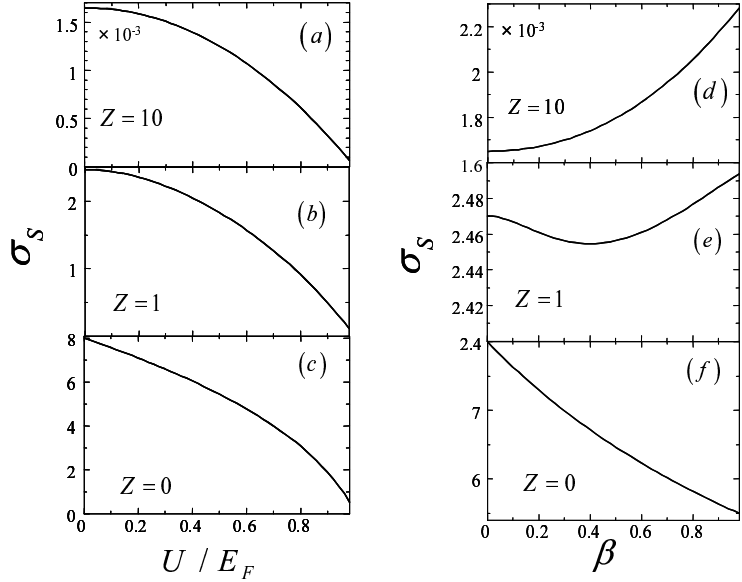}}
\end{center}
\par
\caption{ Tunneling condutance for superconducting states  at zero voltage as a function of the exchange field in F/S junctions (left panels) and RSOC in 2DEG/S junctions  (right panels) with $Z=10$ in (a) and (d), $Z=1$ in (b) and (e), and  $Z=0$ in (c) and (f). }
\label{f1}
\end{figure}

First we study the difference between the effect of the Zeeman splitting and that of Rashba splitting. We plot the tunneling condutance for superconducting states, $\sigma _S$ for F/S junctions in (a)-(c) and for 2DEG/S junctions in (d)-(f) of Fig. \ref{f1} with $Z=10$ in (a) and (d), $Z=1$ in (b) and (e), and  $Z=0$ in (c) and (f). The exchange field suppresses  $\sigma _S$ independently of $Z$ as shown in (a)-(c). This is because the AR probability is reduced by the  exchange field. 
On the other hand the dependence of $\sigma _S$ on $\beta$ at zero voltage is qualitatively different. In (d)-(f) we show the dependence of $\sigma _S$  on $\beta$ at zero voltage for various $Z$. For $Z=10$ it has an exponential dependence on $\beta$ but its magnitude is very small. For $Z=1$ it has a reentrant behavior as a function of $\beta$. For $Z=0$ it decreases linearly as a function of $\beta$. The  AR probability is enhanced by the RSOC at $Z=10$. 

\begin{figure}[htb]
\begin{center}
\scalebox{0.4}{
\includegraphics[width=28.0cm,clip]{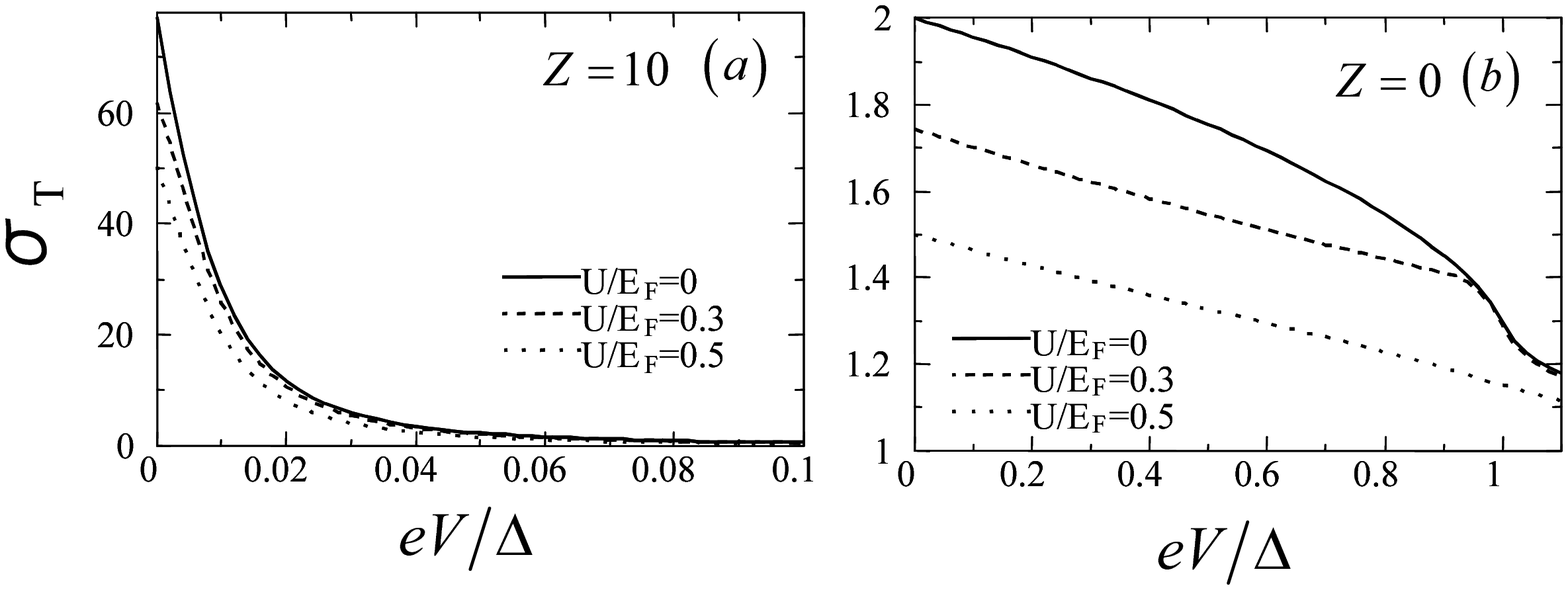}}
\end{center}
\par
\caption{ Normalized tunneling conductance in F/D junctions at $\alpha/\pi=0.25$ with $Z=10$ in (a) and $Z=0$ in (b).}
\label{f2}
\end{figure}

Next we will study the F/D junctions. The normalized tunneling conduntace $\sigma _T$ as a function of bias voltage $V$ is plotted in Fig. \ref{f2} for $\alpha/\pi=0.25$ and various exchange field with $Z=10$ in (a) and $Z=0$ in (b). For $Z=10$ a ZBCP appears due to the formation of the MARS as shown in (a). As the exchange field increases,  $\sigma _T$ is suppressed. Similar plots at $Z=0$ are shown in (b). We can find that $\sigma _T$ decreases with the increase of  the exchange field. 

\section{Diffusive junctions}

 We consider a
junction consisting of normal and superconducting reservoirs connected by a
quasi-one-dimensional diffusive ferromagnet conductor (DF) with a length $L$
much larger than the mean free path. The interface between the DF conductor
and the S electrode has a resistance $R_{b}$ while the DF/N interface has a
resistance $R_{b}^{\prime}$. The positions of the DF/N interface and the
DF/S interface are denoted as $x=0$ and $x=L$, respectively. We model
infinitely narrow insulating barriers by the delta function $%
U(x)=H\delta(x-L)+H^{\prime}\delta(x)$. The resulting transparency of the
junctions $T_{m}$ and $T_{m}^{\prime}$ are given by $T_{m}=4\cos ^{2}\phi
/(4\cos ^{2}\phi +Z^{2})$ and $T_{m}^{\prime}=4\cos ^{2}\phi /(4\cos
^{2}\phi +{Z^{\prime}}^{2})$, where $Z=2H/v_{F}$ and $Z^{\prime}=2H^{%
\prime}/v_{F}$ are dimensionless constants and $\phi $ is the injection
angle measured from the interface normal to the junction and $v_{F}$ is
Fermi velocity. 

We apply the quasiclassical Keldysh formalism in the following calculation
of the tunneling conductance. The 4 $\times $ 4 Green's functions in N, DF
and S are denoted by $\check{G}_{0}(x)$, $\check{G}_{1}(x)$ and $\check{G}%
_{2}(x)$ respectively where the Keldysh component $\hat{K}_{0,1,2}(x)$ is given by $\hat{K%
} _{i}(x)=\hat{R}_{i}(x)\hat{f}_{i}(x)-\hat{f}_{i}(x)\hat{A} _{i}(x)$ with
retarded component $\hat{R}_{i}(x)$, advanced component $\hat{A}_{i}(x)=-%
\hat{R}_{i}^{\ast }(x)$ using distribution function $\hat{f}_{i}(x)
(i=0,1,2) $. In the above, $\hat{R}_{0}(x)$ is expressed by $\hat{R}_{0}(x)=%
\hat{\tau}_{3} $ and $\hat{f}_{0}(x)=f_{l0}+\hat{\tau}_{3}f_{t0}$. $\hat{R}%
_{2}(x)$ is expressed by 
$\hat{R}_{2}(x)=g\hat{\tau}_{3}+f\hat{\tau}_{2}$ with $g=\epsilon /\sqrt{\epsilon ^{2}-\Delta^{2}}$ and $f=\Delta/ \sqrt{%
\Delta^{2}-\epsilon ^{2}}$, where $\hat{\tau}_{2}$ and $\hat{\tau}_{3}$ are
the Pauli matrices, and $\varepsilon$ denotes the quasiparticle energy
measured from the Fermi energy and $\hat{f}_{2}(x)=\mathrm{{tanh}[\epsilon
/(2T)]}$ in thermal equilibrium with temperature $T$. 
We put the electrical potential zero in the S-electrode. In this case the
spatial dependence of $\check{G}_{1}(x)$ in DF is determined by the static
Usadel equation \cite{Usadel},

\begin{equation}
D \frac{\partial }{\partial x} [\check{G}_{1} (x) \frac{\partial \check{G}%
_{1}(x) }{\partial x} ] + i [\check{H},\check{G}_{1}(x)] =0
\end{equation}
with the diffusion constant $D$ in DF. Here $\check{H}$ is given by 
\begin{equation}
\check{H}= \left( 
\begin{array}{cc}
\hat{H}_{0} & 0 \\ 
0 & \hat{H}_{0}%
\end{array}
\right),
\end{equation}
with $\hat{H}_{0}=(\epsilon-(+)h) \hat \tau_{3} $ for majority(minority)
spin where $h$ denotes the exchange field. Note that we assume a weak
ferromagnet and neglect the difference of Fermi velocity between majority
spin and minority spin. 
The Nazarov's generalized boundary condition for $\check{ G}_{1}(x)$ at the
DF/S interface is given by Refs.\cite{TGK,Golubov2}. We also use Nazarov's
generalized boundary condition for $\check{ G}_{1}(x)$ at the DF/N
interface: 
\begin{equation}
\frac{L}{R_{d}} (\check{ G}_{1} \frac{ \partial \check{G}_{1} }{\partial x}
)_{\mid x=0_{+}} =-{R_{b}^{\prime}}^{-1}<B>^{\prime}, 
B= \frac{ 2T_{m}^{\prime}[\check{G}_{0}(0_{-}),\check{G}_{1}(0_{+})] } { 4 +
T_{m}^{\prime}([ \check{G}_{0}(0_{-}),\check{G}_{1}(0_{+})]_{+} - 2) }. \label{Nazarov}
\end{equation}
The average over the various angles of injected particles at the interface
is defined as 
\begin{equation}
<B(\phi)>^{(\prime)} = \frac{\int_{-\pi/2}^{\pi/2} d\phi \cos\phi B(\phi)}{
\int_{-\pi/2}^{\pi/2} d\phi T^{(\prime)}(\phi)\cos\phi}
\end{equation}
with $B(\phi)=B$ and $T^{(\prime)}(\phi)=T_{m}^{(\prime)}$. The resistance
of the interface $R_{b}$ is given by 
\begin{equation}
R_{b}^{(\prime)}=R_{0}^{(\prime)} \frac{2} {\int_{-\pi/2}^{\pi/2} d\phi
T^{(\prime)}(\phi)\cos\phi}.
\end{equation}
Here $R_{0}^{(\prime)}$ is Sharvin resistance, which in three-dimensional
case is given by $R_{0}^{(\prime)-1}=e^{2}k_{F}^2S_c^{(\prime)}/(4\pi^{2} )$%
, where $k_{F}$ is the Fermi wave-vector and $S_c^{(\prime)}$ is the
constriction area. Note that the area $S_c^{(\prime)}$ is in general not
equal to the crossection area $S_d$ of the normal conductor, therefore $%
S_c^{(\prime)}/S_d$ is independent parameter of our theory. This allows to
vary $R_{d}/R_{b}$ independently of $T_{m}$. In real physical situation, the
assumption $S^{(\prime)}_{c}<S_{d}$ means that only a part of the actual
flat DN/S interface (having area $S^{(\prime)}_{c}$) is conducting whether it a single conducting region or a series of such regions. These
conducting regions are not constrictions in a standard sense - we don't
assume the narrowing of the total cross-section, but rather that only the
part of the cross-section is conducting.

The electric current per one spin is expressed using $\check{G}_{1}(x)$ as 
\begin{equation}
I_{el}=\frac{-L}{8eR_{d}}\int_{0}^{\infty }d\epsilon \mathrm{Tr}[\hat{\tau
_{3}}( \check{G}_{1}(x)\frac{\partial \check{G}_{1}(x)}{\partial x})^{K}],
\end{equation}%
where $(\check{G_{1}}(x)\frac{\partial \check{G_{1}}(x)}{\partial x})^{K}$
denotes the Keldysh component of $(\check{G_{1}}(x)\frac{\partial \check{%
G_{1}}(x)}{\partial x})$. %
In the actual calculation it is convenient to use the standard $\theta$%
-parameterization where the retarded Green's fucntion 
$\hat{R}_{1}(x)$ is expressed as 
$\hat{R}_{1}(x)=\hat{\tau}_{3}\cos \theta (x)+\hat{\tau}_{2}\sin \theta (x).$ The parameter $\theta (x)$ is a measure of the proximity effect in DF.

The distribution function $\hat{ f} _{1}(x)$ is given by $\hat{f}%
_{1}(x)=f_{l}(x)+\hat{\tau}_{3}f_{t}(x) $. In the above, $f_{t}(x)$ is the
relevant distribution function which determines the conductance of the
junction we are now concentrating on. From the retarded or advanced
component of the Usadel equation, the spatial dependence of $\theta (x)$ is
determined by the following equation 
\begin{equation}
D\frac{\partial ^{2}}{\partial x^{2}}\theta (x)+2i(\epsilon-(+)h )\sin
[\theta (x)]=0  \label{Usa1}
\end{equation}%
for majority(minority) spin, %
while for the Keldysh component we obtain 
\begin{equation}
D\frac{\partial }{\partial x}[\frac{\partial f_{t}(x)}{\partial x} \mathrm{%
cosh^{2}}\theta _{im}(x)]=0.  \label{Usa2}
\end{equation}%
%
At $x=0$, since $f_{t0}$ is the distribution function in the normal
electrode, it is satisfied with 
\begin{equation}
f_{t0}=\frac{1}{2}\{\tanh [(\epsilon +eV)/(2T)]-\tanh [(\epsilon
-eV)/(2T)]\}.
\end{equation}%
%
Next we focus on the boundary condition at the DF/N interface. Taking the
retarded part of Eq.~(\ref{Nazarov}), we obtain 
\begin{equation}
\frac{L}{R_{d}}\frac{\partial \theta (x)}{\partial x}\mid _{x=0_{+}}=\frac{
<F>^{\prime}}{R_{b}^{\prime}}
F = \frac{2T_{m}^{\prime} \sin\theta_{0} } { (2-T_{m}^{\prime}) +
T_{m}^{\prime}\cos \theta_{0}},
\end{equation}
with $\theta_{0}=\theta(0_{+})$.

On the other hand, from the Keldysh part of Eq.~(\ref{Nazarov}), we obtain 
\begin{equation}
\frac{L}{R_{d}} (\frac{\partial f_{t}}{ \partial x}) \mathrm{{\cosh^{2}}}
\theta_{im}(x) \mid_{x=0_{+}} =-\frac{<I_{b1}>^{\prime}(f_{t0}- f_{t}(0_{+}))%
}{R_{b}^{\prime}},
I_{b1} = \frac{T_{m}^{\prime2}\Lambda_{1}^{\prime} +
2T_{m}^{\prime}(2-T_{m}^{\prime})\mathrm{{Real} \{\cos \theta_{0}\}}} { \mid
(2-T_{m}^{\prime}) + T_{m}^{\prime}\cos\theta_{0} \mid^{2} }
\end{equation}

\begin{equation}
\Lambda_{1}^{\prime}=(1+\mid \cos\theta_{0} \mid^{2} + \mid \sin\theta_{0}
\mid^{2}).
\end{equation}
After some calculations we obtain the following final result for the current

\begin{equation}
I_{el}=\frac{1}{2e}\int_{0}^{\infty }d\epsilon \frac{f_{t0}}{\frac{R_{b}}{
<I_{b0}>}+\frac{R_{d}}{L}\int_{0}^{L}\frac{dx}{\cosh ^{2} \theta _{im}(x)}+%
\frac{R_{b}^{\prime}}{<I_{b1}>^{\prime}}}.
\end{equation}%
Then the differential resistance $R$ per one spin projection at zero
temperature is given by 
\begin{equation}
R=\frac{2R_{b}}{<I_{b0}>}+\frac{2R_{d}}{L}\int_{0}^{L}\frac{dx}{\cosh
^{2}\theta _{im}(x)}+\frac{2R_{b}^{\prime}}{<I_{b1}>^{\prime}}
\end{equation}
with 
\begin{equation}
I_{b0} = \frac{T_{m}^2 \Lambda_{1} + 2T_{m}(2-T_{m}) \Lambda_{2}} {2 \mid
(2-T_{m}) + T_{m}[g \cos\theta_{L} + f \sin\theta_{L} ] \mid^{2} },
\end{equation}
\begin{equation}
\Lambda_{1}=(1+\mid \cos\theta_{L} \mid^{2} + \mid \sin\theta_{L} \mid^{2})
(\mid g \mid^{2} + \mid f \mid^{2} +1)
+ 4\mathrm{Imag}[fg^{*}] \mathrm{Imag}[\cos \theta_{L} \sin\theta_{L}^{*} ],
\end{equation}

\begin{equation}
\Lambda_{2} =\mathrm{{Real} \{ g(\cos \theta_{L} + \cos \theta_{L}^{*}) +
f(\sin \theta_{L} + \sin \theta_{L}^{*}) \}}.
\end{equation}

This is an extended version of the VZK formula\cite{Volkov}.  In the above $\theta
_{im}(x)$ and $\theta _{L}$ denote the imaginary
part of $\theta (x)$ and $\theta (L_{-})$ respectively. Then the total
tunneling conductance in the superconducting state $\sigma _{S}(eV)$ is
given by $\sigma _{S}(eV)=\sum_{\uparrow ,\downarrow }1/R$. The local\
normalized DOS $N(\varepsilon )$ in the DF layer is given by 
\begin{equation}
N(\varepsilon )=\frac{1}{2}\sum_{\uparrow ,\downarrow }{\mathop{\rm Re}%
\nolimits}\cos \theta (x).
\end{equation}

It is important to note that in the present approach, according to the
circuit theory, $R_{d}/R^{(\prime)}_{b}$ can be varied independently of $%
T^{(\prime)}_{m}$, $i.e.$, independently of $Z^{(\prime)}$, since one can
change the magnitude of the constriction area $S_c^{(\prime)}$
independently. In other words, $R_{d}/R_{b}^{(\prime)}$ is no more
proportional to $T^{(\prime)}_{av}(L/l)$, where $T^{(\prime)}_{av}$ is the
averaged transmissivity of the barrier and $l$ is the mean free path in the
diffusive region. Based on this fact, we can choose $R_{d}/R^{(\prime)}_{b}$
and $Z^{(\prime)}$ as independent parameters. 

In the following, we will discuss the normalized tunneling
conductance $\sigma _{T}(eV)=\sigma _{S}(eV)/\sigma _{N}(eV)$ where $\sigma
_{N}(eV)$ is the tunneling conductance in the normal state given by $\sigma
_{N}(eV)=\sigma _{N}=1/(R_{d}+R_{b}+R_{b}^{\prime})$.



Now we study the influence of the resonant proximity effect on
tunneling conductance as well as the DOS in the DF region. The resonant
proximity effect was discussed in Ref. \cite{Yoko2} and can be
characterized as follows. When the proximity effect is weak ($R_{d}/R_{b}\ll
1$), the resonant condition is given by $R_{d}/R_{b}\sim 2h/E_{Th}$ due to the 
exchange splitting of DOS in different spin subbands. When the proximity
effect is strong ($R_{d}/R_{b}\gg 1$), the condition is given by $E_{Th}\sim
h$ and is realized when the length of a ferromagnet is equal to the
coherence length $\xi _{F}=\sqrt{D/h}.$ We choose $R_{d}/R_{b}=1$ as a
typical value to study the weak proximity regime. We also choose $%
R_{d}/R_{b}=5$ to study the strong one.
 We fix $Z=Z^{\prime}=3$
because these parameter don't change the results qualitatively and consider
the case of high barrier at the N/DF interface, $R_d/R^{\prime}_b=0.1$, in
order to enhance the proximity effect.

\begin{figure}[htb]
\begin{center}
\scalebox{0.4}{
\includegraphics[width=28.0cm,clip]{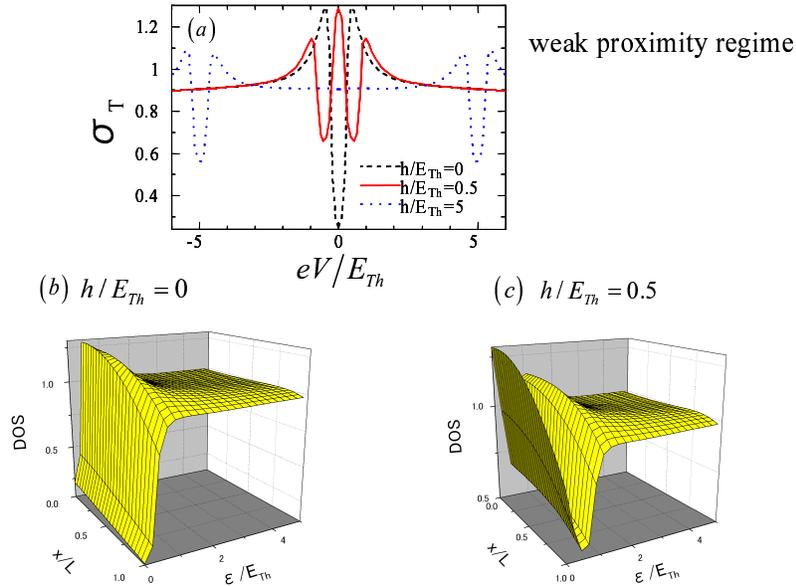}}
\end{center}
\caption{ Normalized tunneling conductance (a) and the DOS ((b) and (c)) with $R_{d}/R_{b}=1$ and $E_{Th}/\Delta =0.01$. (b) $h/E_{Th}=0$ and (c) $h/E_{Th}=0.5$.}
\label{f3}
\end{figure}

Let us first choose the weak
proximity regime and relatively small Thouless energy, $E_{Th}/\Delta =0.01$%
. In this case the resonant condition is satisfied for $h/E_{Th}=0.5$.
In Fig. \ref{f3} we show the tunneling conductance for $R_{d}/R_{b}=1$, $
E_{Th}/\Delta =0.01$  and various $h/E_{Th} $ in (a). 
The ZBCD occur due to the proximity effect for $h=0$. For $h/E_{Th}
=0.5$, the resonant ZBCP appears and split into two peaks or dips at $%
eV\sim \pm h$ with increasing $h/E_{Th} $. The value of the resonant ZBCP
exceeds unity. Note that ZBCP due to the conventional proximity effect in
DN/S junctions is always less than unity \cite%
{Kastalsky,TGK}
and therefore is qualitatively different from the resonant ZBCP in the
 DF/S junctions.

The corresponding normalized DOS of the DF is shown in (b) and (c) of 
Fig. \ref{f3}. Note that in the DN/S junctions, the proximity effect is almost
independent on $Z$ parameter\cite{TGK}. We have checked numerically that this also
holds for the  proximity effect in DF/S junctions. Figure \ref{f3} displays the DOS for $Z=3$, $R_{d}/R_{b}=1$ and $
E_{Th}/\Delta =0.01$ with (b) $h/E_{Th}=0$ and (c) $h/E_{Th}=0.5$
corresponding to the resonant condition. For $h=0$, a sharp dip appears at
zero energy over the whole DF region. For nonzero energy, the DOS is almost
unity and spatially independent. For $h/E_{Th}=0.5$ a zero energy peak
appears in the region of DF near the DF/N interface. This peak is
responsible for the large ZBCP shown in (a). Therefore, the ZBCP in DF/S
junctions has different physical origin compared to the one in DN/S
junctions.

\begin{figure}[htb]
\begin{center}
\scalebox{0.4}{
\includegraphics[width=28.0cm,clip]{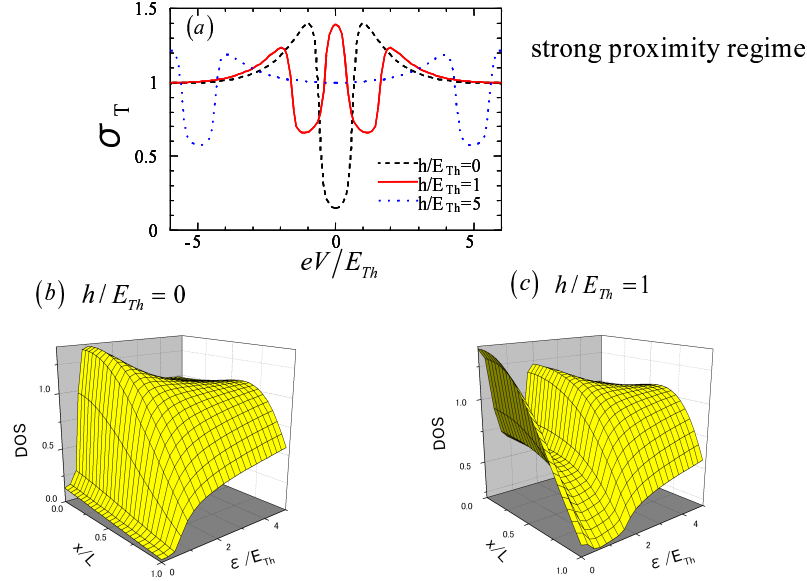}}
\end{center}
\caption{Normalized tunneling conductance (a) and the DOS ((b) and (c)) with  $R_{d}/R_{b}=5$ and $E_{Th}/\Delta =0.01$. (b) $h/E_{Th}=0$ and (c) $h/E_{Th}=1$.}
\label{f4}
\end{figure}

Next we choose the strong proximity regime and relatively small Thouless
energy, $E_{Th}/\Delta =0.01$. In the present case, the resonant ZBCP is
expected for $h/E_{Th}=1$. Figure \ref{f4} displays the tunneling conductance
for $R_{d}/R_{b}=5$ and $%
E_{Th}/\Delta =0.01$ and various $h/E_{Th} $ in (a). 
We can find resonant ZBCP and splitting of the peak as shown in (a). The corresponding DOS is shown in
 (b) $h/E_{Th}=0$ and (c) $h/E_{Th}=1$. For $h=0$, a sharp dip
appears at zero energy. For finite energy the DOS is almost unity and
spatially independent. For $h/E_{Th}=1$ a peak occurs at zero energy in
the range of $x$ near the DF/N interface. We can find a similar structure in
the corresponding conductance as shown in Fig. \ref{f4}(a). The DOS around zero
energy is strongly suppressed at the DF/S interface $(x=L)$ compared to the
one in Fig. \ref{f3}.

Let us proceed to study the $d$-wave junctions both for weak and strong proximity
regimes. In this case, depending on the orientation angle $\alpha $, the
proximity effect is drastically changed: As $\alpha$ increases the proximity effect reduces\cite{Nazarov3,Golubov2}. For $\alpha =0$ we can expect
similar results to those in  the $s$-wave junctions since proximity effect exists
while the MARS is absent. In contrast, the  tunneling
conductance for large $\alpha$ is almost independent of $h/E_{Th} $. Especially, the conductance is
independent of $h$ for $\alpha /\pi =0.25$ due to the complete absence of
the proximity effect. There exist two different origins for ZBCP in
DF/D junctions: the ZBCP by resonant proximity effect peculiar to DF
and the ZBCP by the MARS formed at DF/D interface. When $\alpha $ deviates from 0, the MARS are formed at the interface. 
At the same time, the proximity effect is suppressed due to the 
competition between the proximity effect and the MARS. 
Therefore the MARS provide the dominant contribution to the ZBCP 
compared to the resonant proximity effect in DF, as will be discussed below.

\begin{figure}[htb]
\begin{center}
\scalebox{0.4}{
\includegraphics[width=30.0cm,clip]{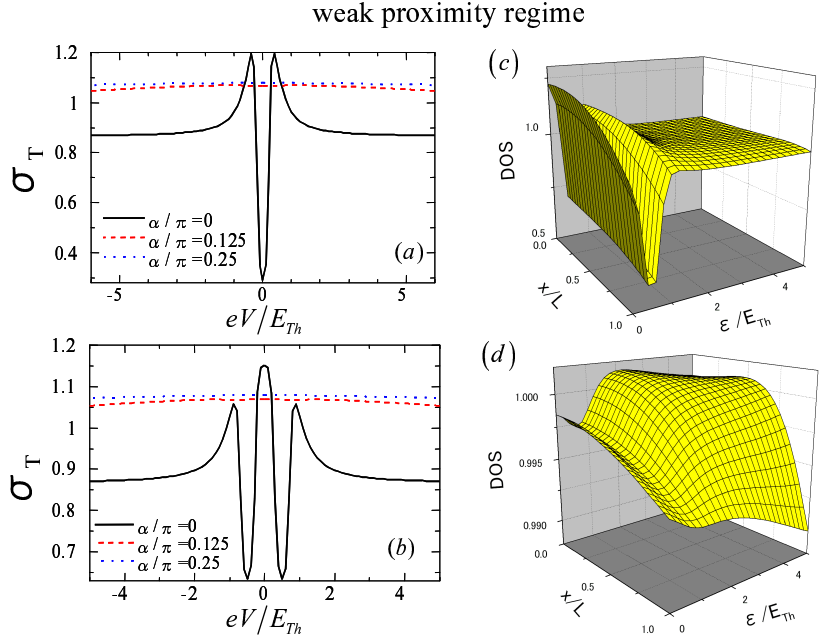}}
\end{center}
\caption{Normalized tunneling conductance ((a) and (b))  and the DOS ((c) and (d)) for $d$-wave
superconductors with  $R_{d}/R_{b}=1$ and $E_{Th}/\Delta =0.01$. (a) $h/E_{Th}=0$, (b) $h/E_{Th}=1$ (c) $h/E_{Th}=1$ and $
\protect\alpha/\protect\pi=0$ and (d) $h/E_{Th}=1$ and $\protect\alpha/\protect\pi=0.125$.}
\label{f5}
\end{figure}

First we choose the weak proximity regime where the resonant condition is $%
h/E_{Th}=0.5$. Figure \ref{f5} displays the tunneling conductance for  $R_d/R_b=1$, $E_{Th}/\Delta=0.01$ and  various $\alpha$
with (a) $h/E_{Th}=0$ and (b) 
$h/E_{Th}=0.5$. For  $h=0$, ZBCD
appears for $\alpha/\pi=0$ due to the proximity effect as in the case of the 
$s$-wave junctions while ZBCP appears  for $\alpha/\pi=0.25$ 
due to the formation of the MARS (Fig. \ref{f5} (a)). For $h/E_{Th}=0.5$, the height of the ZBCP by the
resonant proximity effect for $\alpha=0$ exceeds the one by MARS for $\alpha/\pi=0.25$ 
(Fig. \ref{f5} (b)) in contrast to  the ballistic junctions where the ZBCP for $%
\alpha/\pi=0.25$ is most enhanced\cite{TK95}.

We also study the DOS of the DF for the same parameters as those of   Fig. \ref
{f5}(b)  as shown in (c) $\alpha/\pi=0$ and (d) $\alpha/\pi=0.125$ in Fig. 
\ref{f5}. The line shapes of the LDOS near the DF/S interface are qualitatively similar to the
tunneling conductance. For $\alpha/\pi=0$ a zero-energy peak appears as in the case of $s$%
-wave junctions. With increasing $\alpha$, the DOS around the zero energy is suppressed due to the reduction of the proximity effect. The extreme case is 
$\alpha/\pi=0.25$, where the DOS is always unity since the proximity effect
is completely absent.

\begin{figure}[htb]
\begin{center}
\scalebox{0.4}{
\includegraphics[width=30.0cm,clip]{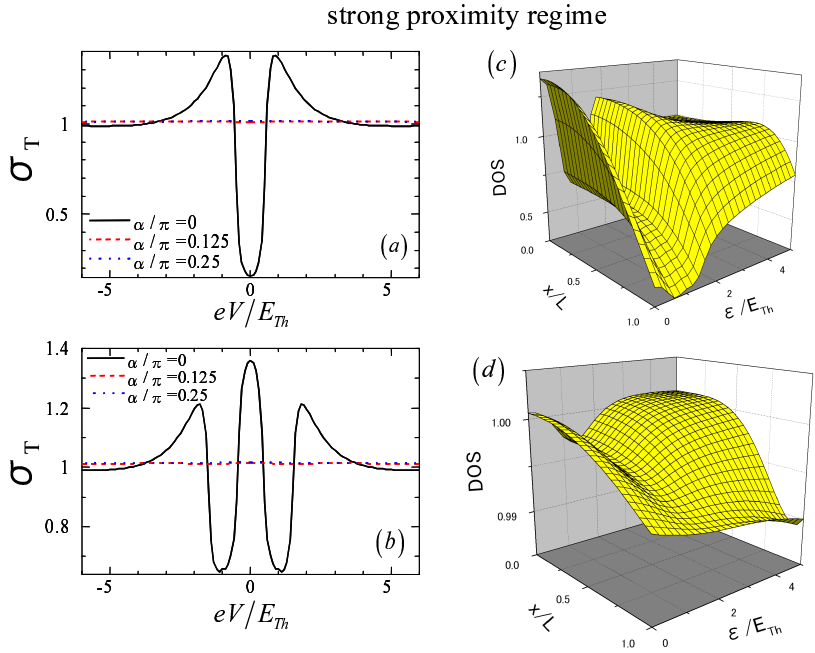}}
\end{center}
\caption{Normalized tunneling conductance ((a) and (b))  and the DOS ((c) and (d)) for $d$-wave
superconductors with  $R_{d}/R_{b}=5$ and $E_{Th}/\Delta =0.01$. (a) $h/E_{Th}=0$, (b) $h/E_{Th}=1$ (c) $h/E_{Th}=1$ and $
\protect\alpha/\protect\pi=0$ and (d) $h/E_{Th}=1$ and $\protect\alpha/\protect\pi=0.125$.}
\label{f6}
\end{figure}

Next we look at the junctions for the strong proximity regime. In Fig. \ref{f6}
 we show the tunneling conductance for  $R_d/R_b=5$, $E_{Th}/\Delta=0.01$
and  various $\alpha$ with (a)
 $h/E_{Th}=0$ and  (b) $h/E_{Th}=1$. In this case we also find the ZBCP for $\alpha=0$ by the
resonant proximity effect. 
The height of the ZBCP is suppressed as $\alpha$ increases as shown in Figs. \ref{f6}(b). 

The corresponding DOS of the DF to the case of (b) in Fig. \ref{f6} is shown in  Fig. \ref{f6} with
(c) $\alpha/\pi=0$ and (d) $\alpha/\pi=0.125$. 
 For $\alpha=0$ the structure of the DOS at $x=0$ reflects the line shapes of the tunneling conductance. With increasing $\alpha$ the zero energy peak of the DOS becomes 
suppressed.   The DOS at the DF/S interface ($x=L$) are drastically
suppressed  compared to those in Fig. \ref{f5}.

\clearpage
\section{Conclusions}

In this article we  studied the tunneling conductance in F/S, F/D and 2DEG/S junctions in ballistic regime.  We  extended the  BTK formula and calculated the tunneling conductance of the junctions. We clarified the following point:

1. The exchange field always suppresses the conductance in F/S and F/D juntions.

2. In 2DEG/S jucntions, for low insulating barrier  the RSOC suppresses the tunneling conductance  while for high insulating barrier it can slightly  enhance the tunneling conductance. We also found a reentrant behavior of the conductance at zero voltage as a function of RSOC for intermediate insulating barrier strength. The results give  the possibility to control the AR probability by a gate voltage. We believe that the obtained results are useful for a better understanding of related experiments of   mesoscopic F/S and 2DEG/S junctions.

 In the latter of the present paper, a detailed theoretical study of the tunneling
conductance and the density of states in normal metal / diffusive
ferromagnet / $s$- and $d$-wave superconductor junctions is presented. We
 clarified that the resonant proximity effect strongly influences the
tunneling conductance and the density of states. There are several points
which have been clarified in this paper:

3. For $s$-wave junctions, due to the resonant proximity effect, a sharp
ZBCP appears for small $E_{Th}$
. We showed that the mechanism of the ZBCP in DF/S junctions is essentially
different from that in DN/S junctions and is due to the strong enhancement of
DOS at a certain value of the exchange field. As a result, the magnitude of 
the ZBCP in DF/S junctions  can exceed unity in contrast to  that in DN/S junctions.

4. For $d$-wave junctions at $\alpha =0$, similar
to the s-wave case, the sharp ZBCP is formed when the resonant condition is
satisfied. At finite misorientation angle $\alpha$, the MARS contribute to
the conductance when $R_{d}/R_{b}\ll 1$ and $Z \gg 1$. With the increase of $\alpha $ the contribution of the resonant proximity
effect becomes smaller while the MARS dominate the conductance. As a result,
for sufficiently large $\alpha$ ZBCP exists
independently of whether the resonant condition  is satisfied or not.
In the opposite case of the weak barrier, $R_{d}/R_{b}\gg 1$, the
contribution of MARS is negligible and ZBCP appears only when the resonant
condition is satisfied. 

An interesting problem is a calculation of the tunneling conductance in
normal metal / diffusive ferromagnet / $p$-wave superconductor junctions
because interesting phenomena were predicted in diffusive normal metal / $p$%
-wave superconductor junctions\cite{p-wave}. We will perform it in the near
future.



\section*{Acknowledgements}
The authors appreciate useful and fruitful discussions with J. Inoue, Yu. Nazarov and  A. Golubov. 
This work was supported by
NAREGI Nanoscience Project, the Ministry of Education, Culture,
Sports, Science and Technology, Japan, the Core Research for Evolutional
Science and Technology (CREST) of the Japan Science and Technology
Corporation (JST) and a Grant-in-Aid for the 21st Century COE "Frontiers of
Computational Science". 
The computational aspect of this work has been
performed at the Research Center for Computational Science, Okazaki National
Research Institutes and the facilities of the Supercomputer Center,
Institute for Solid State Physics, University of Tokyo and the Computer Center.


\begin{thebibliography}{100}
\bibitem{Andreev} A.F. Andreev, Sov. Phys. JETP \textbf{19}, 1228 (1964).

\bibitem{BTK} G.E. Blonder, M. Tinkham, and T.M. Klapwijk, Phys. Rev. B 
\textbf{25}, 4515 (1982).

\bibitem{TK95} Y. Tanaka and S. Kashiwaya, Phys. Rev. Lett. \textbf{74},
3451 (1995); S. Kashiwaya, Y. Tanaka, M. Koyanagi and K. Kajimura, Phys.
Rev. B \textbf{53}, 2667 (1996); Y. Tanuma, Y. Tanaka, and S. Kashiwaya
Phys. Rev. B \textbf{64}, 214519 (2001).

\bibitem{FS} T. Hirai, Y. Tanaka, N. Yoshida, Y. Asano, J. Inoue, and S.
Kashiwaya Phys. Rev. B \textbf{67}, 174501 (2003); N. Yoshida, Y. Tanaka, J.
Inoue, and S. Kashiwaya, J. Phys. Soc. Jpn. \textbf{68}, 1071 (1999); S.
Kashiwaya, Y. Tanaka, N. Yoshida, and M.R. Beasley, Phys. Rev. B \textbf{60}%
, 3572 (1999); I. Zutic and O.T. Valls, Phys. Rev. B \textbf{60}, 6320
(1999); \textbf{61}, 1555 (2000); N. Stefanakis, Phys. Rev. B \textbf{64},
224502 (2001); J. Phys. Condens. Matter \textbf{13}, 3643 (2001).

\bibitem{Tedrow} P. M. Tedrow and R. Meservey, Phys. Rev. Lett. \textbf{26},
192 (1971); Phys. Rev. B \textbf{7}, 318 (1973); R. Meservey and P. M.
Tedrow, Phys. Rep. \textbf{238}, 173 (1994).

\bibitem{Upadhyay} S. K. Upadhyay, A. Palanisami, R. N. Louie, and R. A.
Buhrman Phys. Rev. Lett. \textbf{81}, 3247 (1998).

\bibitem{Soulen} R. J. Soulen Jr., J. M. Byers, M. S. Osofsky, B. Nadgorny,
T. Ambrose, S. F. Cheng, P. R. Broussard, C. T. Tanaka, J. Nowak, J. S.
Moodera, A. Barry, J. M. D. Coey Science \textbf{282}, 85 (1998)


\bibitem{de Jong} M.J.M. de Jong and C.W.J. Beenakker, Phys. Rev. Lett. \textbf{74}, 1657 (1995). 

\bibitem{Hirsch} J. E. Hirsch, Phys. Rev. Lett. \textbf{83} (1999) 1834. 

\bibitem{Streda} P. Streda,  P. Seba, Phys. Rev. Lett. \textbf{90} (2003) 256601.

\bibitem{Schliemann} J. Schliemann,  D. Loss, Phys. Rev. B \textbf{68} (2003) 165311.

\bibitem{Sinova} J. Sinova, D. Culcer, Q. Niu, N. A. Sinitsyn, T. Jungwirth,  A. H. MacDonald, Phys. Rev. Lett. \textbf{92} (2004) 126603.

\bibitem{Rashba} E. I. Rashba, Fiz. Tverd. Tela (Leningrad) \textbf{2} (1960) 1224;
[Sov. Phys. Solid State \textbf{2}(1960) 1109 ];
Yu. A. Bychkov,  E. I. Rashba, J. Phys. C \textbf{17} (1984) 6039.

\bibitem{Datta} S. Datta,  B. Das, Appl. Phys. Lett. \textbf{56} (1990) 665.

\bibitem{Molenkamp} L. W. Molenkamp, G. Schmidt,  G. E. W. Bauer, Phys. Rev. B \textbf{64} (2001) 121202.

\bibitem{Edelstein} V. M. Edelstein, Solid State Commun. \textbf{73} (1990) 233. 
\bibitem{Inoue} J. Inoue, G. E. W. Bauer,  L. W. Molenkamp, Phys. Rev. B \textbf{67} (2003) 033104; \textbf{70} (2004) 041303.


\bibitem{Hekking} F. W. J. Hekking and Yu. V. Nazarov, Phys. Rev. Lett. 
\textbf{71}, 1625 (1993). 

\bibitem{Giazotto} F. Giazotto, P. Pingue, F. Beltram, M. Lazzarino, D.
Orani, S. Rubini, and A. Franciosi, Phys. Rev. Lett. \textbf{87}, 216808
(2001).

\bibitem{Klapwijk} T.M. Klapwijk, Physica B \textbf{197}, 481 (1994).

\bibitem{Kastalsky} A. Kastalsky, A.W. Kleinsasser, L.H. Greene, R. Bhat,
F.P. Milliken, J.P. Harbison, Phys. Rev. Lett. \textbf{67}, 3026 (1991).

\bibitem{Nguyen} C. Nguyen, H. Kroemer and E.L. Hu, Phys. Rev. Lett. \textbf{%
69}, 2847 (1992).

\bibitem{Wees} B.J. van Wees, P. de Vries, P. Magnee, and T.M. Klapwijk,
Phys. Rev. Lett. \textbf{69}, 510 (1992).

\bibitem{Nitta} J. Nitta, T. Akazaki and H. Takayanagi, Phys. Rev. B \textbf{%
49} 3659 (1994).

\bibitem{Bakker} S.J.M. Bakker, E. van der Drift, T.M. Klapwijk, H.M.
Jaeger, and S. Radelaar, Phys. Rev. B \textbf{49}, 13275 (1994).

\bibitem{Xiong} P. Xiong, G. Xiao and R.B. Laibowitz, Phys. Rev. Lett. 
\textbf{71}, 1907 (1993).

\bibitem{Magnee} P.H.C. Magnee, N. van der Post, P.H.M. Kooistra, B.J. van
Wees, and T.M. Klapwijk, Phys. Rev. B \textbf{50}, 4594 (1994).

\bibitem{Kutch} J. Kutchinsky, R. Taboryski, T. Clausen, C. B. Sorensen, A.
Kristensen, P. E. Lindelof, J. Bindslev Hansen, C. Schelde Jacobsen, and J.
L. Skov, Phys.Rev. Lett. \textbf{78} ,931 (1997).

\bibitem{Poirier} W. Poirier, D. Mailly, and M. Sanquer, Phys. Rev. Lett. 
\textbf{79}, 2105 (1997).

%


\bibitem{Eilenberger} G.Eilenberger,Z.Phys.\textbf{214},195 (1968)

\bibitem{Eliashberg} G.M. Eliashberg, Sov. Phys. JETP \textbf{34}, 668 (1971)

\bibitem{Larkin} A.I. Larkin and Yu. V. Ovchinnikov, Sov. Phys. JETP \textbf{%
\ 41}, 960 (1975)

\bibitem{Usadel} K.D. Usadel Phys. Rev. Lett. \textbf{25}, 507 (1970).

\bibitem{Volkov} A.F. Volkov, A.V. Zaitsev and T.M. Klapwijk, Physica C 
\textbf{210}, 21 (1993).

\bibitem{KL} M.Yu. Kupriyanov and V. F. Lukichev, Zh. Exp. Teor. Fiz. 
\textbf{94} (1988) 139 [Sov. Phys. JETP \textbf{67}, (1988) 1163].

\bibitem{Nazarov1} Yu. V. Nazarov, Phys. Rev. Lett. \textbf{73}, 1420 (1994).

\bibitem{Yip} S. Yip, Phys. Rev. B \textbf{52}, 3087 (1995).


\bibitem{Stoof} Yu. V. Nazarov and T. H. Stoof, Phys. Rev. Lett. \textbf{76}%
, 823 (1996); T. H. Stoof and Yu. V. Nazarov, Phys. Rev. B \textbf{53},
14496 (1996).

\bibitem{Reentrance} A. F. Volkov, N. Allsopp, and C. J. Lambert, J. Phys.
Cond. Mat. \textbf{8}, L45 (1996); A. F. Volkov and H. Takayanagi, Phys.
Rev. B \textbf{56}, 11184 (1997).

\bibitem{Golubov} A.A. Golubov, F.K. Wilhelm, and A.D. Zaikin, Phys. Rev. B 
\textbf{55}, 1123 (1997).

\bibitem{Takayanagi} A.F. Volkov and H. Takayanagi, Phys. Rev. B \textbf{56}%
, 11184 (1997).

\bibitem{Bezuglyi} E. V. Bezuglyi, E. N. Bratus', V. S. Shumeiko, G. Wendin
and H. Takayanagi, Phys. Rev. B \textbf{62}, 14439 (2000).

\bibitem{Seviour} R. Seviour and A. F. Volkov, Phys. Rev. B \textbf{61},
R9273 (2000).

\bibitem{Belzig} W. Belzig, F. K. Wilhelm, C. Bruder, \textit{et al}.,
Superlattices and Microstructures \textbf{25}, 1251 (1999).

%
%

\bibitem{TGK} Y. Tanaka, A. A. Golubov and S. Kashiwaya, Phys. Rev. B 
\textbf{68} 054513 (2003).

\bibitem{Nazarov2} Yu. V. Nazarov, Superlattices and Microstructuctures 
\textbf{25}, 1221 (1999), cond-mat/9811155.








\bibitem{Buch} L.J. Buchholtz and G. Zwicknagl, Phys. Rev. B \textbf{23}
5788 (1981); C. Bruder, Phys. Rev. B \textbf{41}, 4017 (1990); C.R. Hu,
Phys. Rev. Lett. \textbf{72}, 1526 (1994).

\bibitem{Kashi00} S. Kashiwaya and Y. Tanaka, Rep. Prog. Phys. \textbf{63},
1641 (2000) and references therein. 
%
%
%

\bibitem{Experiments} J. Geerk, X.X. Xi, and G. Linker: Z. Phys. B. \textbf{%
73},(1988) 329; S. Kashiwaya, Y. Tanaka, M. Koyanagi, H. Takashima, and K.
Kajimura, Phys. Rev. B \textbf{51} 1350 (1995); L. Alff, H. Takashima, S.
Kashiwaya, N. Terada, H. Ihara, Y. Tanaka, M. Koyanagi, and K. Kajimura,
Phys. Rev. B {55}, 14757 (1997); M. Covington, M. Aprili, E. Paraoanu, L.H.
Greene, F. Xu, J. Zhu, and C.A. Mirkin, Phys. Rev. Lett. \textbf{79}, 277
(1997); J. Y. T. Wei, N.-C. Yeh, D. F. Garrigus and M. Strasik: Phys. Rev.
Lett. \textbf{81}, (1998) 2542; I. Iguchi, W. Wang, M. Yamazaki, Y. Tanaka,
and S. Kashiwaya: Phys. Rev. B \textbf{62}, (2000) R6131; F. Laube, G. Goll,
H.v. L\"{o}hneysen, M. Fogelstr\"{o}m, and F. Lichtenberg, Phys. Rev. Lett. 
\textbf{84}, 1595 (2000); Z.Q. Mao, K.D. Nelson, R. Jin, Y. Liu, and Y.
Maeno, Phys. Rev. Lett. \textbf{87}, 037003 (2001); Ch. W\"{a}lti, H.R. Ott,
Z. Fisk, and J.L. Smith, Phys. Rev. Lett. \textbf{84}, 5616 (2000); H.
Aubin, L. H. Greene, Sha Jian and D. G. Hinks, Phys. Rev. Lett. \textbf{89},
177001 (2002); Z. Q. Mao, M. M. Rosario, K. D. Nelson, K. Wu, I. G. Deac, P.
Schiffer, Y. Liu, T. He, K. A. Regan, and R. J. Cava Phys. Rev. B \textbf{67}%
, 094502 (2003); A. Sharoni, O. Millo, A. Kohen, Y. Dagan, R. Beck, G.
Deutscher, and G. Koren Phys. Rev. B \textbf{65}, 134526 (2002); A. Kohen,
G. Leibovitch, and G. Deutscher Phys. Rev. Lett. \textbf{90}, 207005 (2003).

\bibitem{Nazarov3} Y. Tanaka, Y.V. Nazarov and S. Kashiwaya, Phys. Rev.
Lett. \textbf{90} 167003 (2003).

\bibitem{Golubov2} Y. Tanaka, Y. Nazarov A. Golubov and S. Kashiwaya, Phys.
Rev. B \textbf{69} 144519 (2004).

\bibitem{p-wave} Y. Tanaka and S. Kashiwaya, Phys. Rev. B \textbf{70} 012507
(2004); Y. Tanaka, S. Kashiwaya, and T. Yokoyama Phys. Rev. B \textbf{71},
094513 (2005).

\bibitem{Yip2} S. Yip, Phys. Rev. B \textbf{52}, 15504 (1995).

\bibitem{Yoko} T. Yokoyama, Y. Tanaka, A. A. Golubov, J. Inoue, and Y.
Asano, Phys. Rev. B \textbf{71}, 094506 (2005).

\bibitem{Buzdin1982} A.I. Buzdin, L.N. Bulaevskii, and S.V. Panyukov, Pis'ma
Zh. Eksp. Teor. Phys. 35, 147, (1982) [JETP Lett. 35, 178 (1982)].

\bibitem{Buzdin1991} A.I. Buzdin and M.Yu. Kupriyanov,, Pis'ma Zh. Eksp.
Teor. Phys. 53, 308 (1991) [JETP Lett. 53, 321 (1991)].

\bibitem{Demler} E. A. Demler, G. B. Arnold, and M. R. Beasley, Phys. Rev. B 
\textbf{55}, 15 174 (1997).

\bibitem{Ryazanov} V. V. Ryazanov, V. A. Oboznov, A. Yu. Rusanov, A. V.
Veretennikov, A. A. Golubov, and J. Aarts, Phys. Rev. Lett. \textbf{86},
2427 (2001); V. V. Ryazanov, V. A. Oboznov, A. V. Veretennikov, and A. Yu.
Rusanov, Phys. Rev. B \textbf{65}, 020501(R) (2001); S. M. Frolov, D. J. Van
Harlingen, V. A. Oboznov, V. V. Bolginov, and V. V. Ryazanov, Phys. Rev. B 
\textbf{70}, 144505.

\bibitem{Kontos1} T. Kontos, M. Aprili, J. Lesueur, F. Genet, B.
Stephanidis, and R. Boursier, Phys. Rev. Lett. \textbf{89}, 137007 (2002).

\bibitem{Blum} Y. Blum, A. Tsukernik, M. Karpovski, and A. Palevski, Phys.
Rev. Lett. \textbf{89}, 187004 (2002).

\bibitem{Sellier} H. Sellier, C. Baraduc, F. Lefloch, and R. Calemczuk,
Phys. Rev. B \textbf{68}, 054531 (2003).

\bibitem{Strunk} A. Bauer, J. Bentner, M. Aprili, M. L. Della Rocca, M.
Reinwald, W. Wegscheider, and C. Strunk, Phys. Rev. Lett.\textbf{\ 92},
217001 (2004).

\bibitem{Radovic} Z. Radovic, M. Ledvij, Lj. Dobrosavljevi-Gruji, A. I.
Buzdin, and J. R. Clem, Phys. Rev. B \textbf{44}, 759 (1991).

\bibitem{Tagirov} L. R. Tagirov, Phys. Rev. Lett. \textbf{83}, 2058 (1999).

\bibitem{Fominov} Ya. V. Fominov, N. M. Chtchelkatchev, and A. A. Golubov,
Pis'ma Zh. Eksp. Teor. Fiz. \textbf{74}, 101 (2001) [JETP Lett. \textbf{74},
96 (2001)]; Ya. V. Fominov, N. M. Chtchelkatchev, and A. A. Golubov, Phys.
Rev. B \textbf{66}, 014507 (2002).

\bibitem{Rusanov} A. Rusanov, R. Boogaard, M. Hesselberth, H. Sellier, and
J. Aarts, Physica C \textbf{36}9, 300 (2002).

\bibitem{Ryazanov1} V.V.Ryazanov, V.A.Oboznov, A.S.Prokofiev, S.V.Dubonos,
JETP Lett. \textbf{77}, 39 (2003).

\bibitem{Kadigrobov} A. Kadigrobov, R. I. Shekhter, M. Jonson and Z. G.
Ivanov, Phys. Rev. B \textbf{60}, 14593 (1999).

\bibitem{Seviour2} R. Seviour, C. J. Lambert, and A. F. Volkov , Phys. Rev.
B \textbf{59}, 6031 (1999).

\bibitem{Leadbeater} M. Leadbeater, C. J. Lambert, K. E. Nagaev, R. Raimondi
and A. F. Volkov, Phys. Rev. B \textbf{59}, 12264 (1999).

\bibitem{Bergeret} F. S. Bergeret, K. B. Efetov, and A. I. Larkin, Phys.
Rev. B \textbf{62}, 11872 (2000); F. S. Bergeret, A. F. Volkov, and K. B.
Efetov, Phys. Rev. Lett. \textbf{86}, 4096 (2001).

\bibitem{Kadigrobov2} A. Kadigrobov, R. I. Shekhter, and M. Jonson,
Europhys. Lett. \textbf{54}, 394 (2001).

\bibitem{Buzdin} A. Buzdin, Phys. Rev. B \textbf{62}, 11 377 (2000).

\bibitem{Zareyan} M. Zareyan, W. Belzig, and Yu. V. Nazarov, Phys. Rev.
Lett. \textbf{86}, 308 (2001); Phys. Rev. B \textbf{65}, 184505 (2002).

\bibitem{Baladie} A. I. Baladie and A. Buzdin, Phys. Rev. B \textbf{64},
224514 (2001).

\bibitem{Bergeret2} F. S. Bergeret, A. F. Volkov, and K. B. Efetov, Phys.
Rev. B \textbf{65}, 134505 (2002).

\bibitem{Golubov3} A. A. Golubov, M. Yu. Kupriyanov, and Ya. V. Fominov,
JETP Lett. \textbf{75}, 223 (2002).

\bibitem{Krivoruchko} V. N. Krivoruchko and E. A. Koshina, Phys. Rev. B 
\textbf{66}, 014521 (2002).

\bibitem{Kontos} T. Kontos, M. Aprili, J. Lesueur, and X. Grison, Phys. Rev.
Lett. \textbf{86}, 304 (2001); T. Kontos, M. Aprili, J. Lesueur, X. Grison,
and L. Dumoulin, Phys. Rev. Lett. \textbf{93}, 137001 (2004). 

\bibitem{Yoko2} T. Yokoyama, Y. Tanaka, and A. A. Golubov, Phys. Rev. B 
\textbf{72}, 052512 (2005). 



\end{thebibliography}
\end{document}